\documentstyle[preprint,eqsecnum,aps]{revtex}
\begin{document}
\draft
\tighten
%\preprint{gr-qc/9707017}

\title{REGULARIZATION OF THE TEUKOLSKY EQUATION FOR ROTATING
BLACK HOLES}
\author{Manuela CAMPANELLI\thanks{Electronic Address:
manuela@mail.physics.utah.edu} and
Carlos O. LOUSTO\thanks{Electronic Address:
lousto@mail.physics.utah.edu}}
\address{Department of Physics, University of Utah\\
201 JBF, Salt Lake City, UT 84112, USA}
\date{\today}
\maketitle

\begin{abstract}
We show that the radial Teukolsky equation (in the frequency domain)
with sources that extend to infinity has well-behaved solutions.
To prove that, we follow Poisson approach to regularize the non-rotating
hole, and extend it to the rotating case. To do so we use the
Chandrasekhar transformation among
the Teukolsky and Regge-Wheeler-like equations, and express the integrals
over the source in terms of solutions to the homogeneous 
Regge-Wheeler-like equation, to finally regularize the resulting
integral. We then discuss the applicability of these results.
\end{abstract}
%\narrowtext\twocolumn

\section{INTRODUCTION}

With the approach to the beginning of a new era in
gravitational wave detectors, a renewed interest in the theoretical
study of astrophysical sources of gravitational radiation arouse.
Several groups of research are now reviewing computations carried
out twenty or more years ago. One of the most remarkable ``old'' results,
is the description of the gravitational perturbations about black
holes. In particular, for the most plausible astrophysical scenario,
i.e. the Kerr (rotating) hole, perturbations are compactly described
by the Teukolsky equation\cite{T72}. Among the new theoretical results,
stands out the close limit approximation\cite{PP94}, which approximates
the collision of two black holes by a single perturbed one. The subsequent
extensions to second order perturbations\cite{GNPP96}, and to moving
(head-on) holes\cite{BAABPPS97}, confirmed the success of the approximation.
Also recently\cite{LP97a,LP97b}, the collision of a small and a big
black hole by perturbative methods have been revisited in order to
incorporate non-vanishing initial data. In both approaches only perturbations
around Schwarzschild (non-rotating) black holes have been studied by
use of the Zerilli-Moncrief equation: A totally well-defined description.
The success of these techniques encourages now the extention of them to a
more realistic rotating background. In the way to obtain such generalization
lie some analytic and conceptual problems. One of them, the regularization
of the Teukolsky equation, have been recently clarified by
Poisson\cite{P97}. Apparently, it was first stressed by Detweiler and
Szedenits\cite{DS79}, that when one tries to find a solution (by
straightforward application of the Green method) to
the radial Teukolsky equation with a source term that extends to
infinity, one obtains a divergent result. Not much later,
Sasaki and Nakamura\cite{SN82} realized that applying a Chandrasekhar
transformation to the radial non-homogeneous (including the source
term) Teukolsky equation 
one could end up with a Regge-Wheeler-like equation (with
a potential term that reduces to the Regge-Wheeler potential in
the $a\to0$ limit), that had a real, short range potential (as opposed to
the original Teukolsky one), and generated sensible finite results
for extending-to-infinity sources.
In this work we generalize Poisson regularization procedure (that he
applied to the non-rotating version of the Teukolsky equation\cite{BP73}),
to the rotating case, which is the relevant case for further applications,
since otherwise one would preferably use the Zerilli-Moncrief formalism.

The paper is organized as follows: In Sec.\ II we review the Teukolsky
equation in its time and frequency domains as well as the asymptotic
behavior of its homogeneous solutions. Section\ III is devoted to recall the main
results of the Chandrasekhar transformation that maps solutions to the
radial Teukolsky equation into solutions to a Regge-Wheeler-like equation.
In Sec.\ IV we formally follow Poisson regularization procedure for
the $a\not=0$ case and in Sec.\ V we explicitly apply it to the problem
of a particle infalling along the symmetry axis of a Kerr hole
$(\theta=0)$. Finally, a brief discussion of the applicability of
these results is given.

Notation and conventions: 
The sign conventions for the metric and curvature tensors are those 
of Misner, Thorne and Wheeler \cite{MTW73}. 
That is, the metric signature is $(-~+~+~+)$; the Riemann tensor is defined
by the equation $R_{~\beta\gamma\delta}^{\alpha}=
\Gamma_{~\beta\gamma ,\delta}^{\alpha}-\Gamma_{~\beta\delta ,\gamma}^{\alpha}
+\Gamma_{~\beta\gamma}^{\sigma}\Gamma_{~\delta\sigma}^{\alpha}-
\Gamma_{~\beta\delta}^{\sigma}\Gamma_{~\gamma\sigma}^{\alpha}$ and the  
Ricci tensor by equation $R_{\mu\nu}=R^{\alpha}_{~\mu\alpha\nu}$.
The tortoise coordinate $r^*$ is defined so that $d/dr^*=f(r)d/dr$,
where $f(r)$ is specified in Sec.\ II.A.
A prime indicates differentiation with
respect to the radial Boyer-Lindquist coordinate $r$.
An overbar represents complex conjugation.

\section{THE TEUKOLSKY EQUATION}

In 1972 Teukolsky\cite{T73}, using the Newman--Penrose formalism,
succeeded in desentangling the perturbations of the Kerr metric, 
and wrote a master equation for fields of arbitrary
spin ($s=0$ for the scalar field, $s=\pm 1$ for the electromagnetic field,
$s=\pm 2$ for the gravitational field, etc).
In Boyer-Lindquist coordinates $(t,r,\theta,\varphi)$
this equation, for pure ``outgoing'' ($s=-2$) gravitational 
perturbations, takes the form
\begin{eqnarray}
\label{master}
&&
\Biggr\{\left[a^2\sin^2\theta-\frac{(r^2 + a^2)^2}{\Delta}\right]
\partial_{tt}-
\frac{4 M a r}{\Delta}\partial_{t\varphi}
+ 4\left[r+ia\cos\theta-\frac{M(r^2-a^2)}{\Delta}\right]\partial_t
\nonumber\\
&&+\,\Delta^{2}\partial_r\left(\Delta^{-1}\partial_r\right)
+\frac{1}{\sin\theta}\partial_\theta\left(\sin\theta\partial_\theta\right)
+\left[\frac{1}{\sin^2\theta}-\frac{a^2}{\Delta}\right]
\partial_{\varphi\varphi}\\
&&-\, 4 \left[\frac{a (r-M)}{\Delta} + \frac{i \cos\theta}{\sin^2\theta}
\right] \partial_\varphi
-\left(4 \cot^2\theta +2 \right)\Biggr\}\Psi=4\pi\Sigma T,\nonumber
\end{eqnarray}
where $M$ is the mass of the black hole, $a$ its angular momentum per
unit mass, $\Sigma\equiv r^2+a^2\cos^2\theta$, and $\Delta\equiv r^2-2Mr+a^2$.

The field $\Psi=\Psi(t,r,\theta,\varphi)$ is the outgoing radiative 
part of the Weyl tensor
\begin{equation}\label{psi}
\Psi=-\rho^{-4}C_{\alpha\beta\gamma\delta}n^\alpha\bar{m}^\beta
n^\gamma\bar{m}^\delta ,
\end{equation}
i. e. one of the two gauge invariant components of the perturbed Weyl 
tensor along the Kinnersley tetrad vectors $n^\alpha$ and $\bar{m}^\alpha$, 
with $\rho$ being a spin coefficient. The explicit expression of the tetrad
vectors and of the spin coefficients are given in Table\ \ref{tabla1},
and the form of the source term, $T=T(t,r,\theta,\varphi)$, in 
equation\ (\ref{master}) is given in Section V, Eq. \ (\ref{fuente}).

Equation\ (\ref{master}) can be separated by mode decomposition of
field and source term in the following way
\begin{equation}\label{separatedpsi}
\Psi=\int{d\omega e^{-i\omega t}\left[\sum_{l,m}R_{lm\omega}(r)
S_{-2,l,m}^{\omega}(\theta)e^{im\varphi}\right]}~,
\end{equation}  
\begin{equation}\label{separatedsource}
4\pi\Sigma T=\int{d\omega e^{-i\omega t}\left[\sum_{l,m}
T_{lm\omega}(r)S_{-2,l,m}^{\omega}(\theta)e^{im\varphi}\right]}~.
\end{equation}

The radial functions $R_{lm\omega}(r)$ satisfy the following equation
\cite{C75}
\begin{equation}
\label{radial}
\left[\Delta^2{d\over dr}\left({1\over\Delta}{d\over dr}\right)
+U(r)\right]R_{lm\omega}(r)=T_{lm\omega}(r),
\end{equation}
with a complex effective potential of the form
\begin{equation}
\label{teukpotential}
U(r)={1\over\Delta}[\omega^2\sigma^2+2i\omega\sigma\Delta'-
(4i\omega\sigma'+\lambda)\Delta],
\end{equation}
where $\sigma\equiv (r^2+a^2)-am/\omega$,
$\lambda\equiv E-2+\omega^2(a^2-am/\omega)$.
$E$ is a separation constant\cite{PT73}, that in the case $a=0$ is given
by $l(l+1)$ and the source term is given in Section V, 
Eq. \ (\ref{radialsource}).

The angular functions $S_{-2,l,m}^{\omega}(\theta)$ are called spin-weighted 
spheroidal harmonics and satisfy an eigenvalue equation of the form
\begin{eqnarray}\label{angular}
\Biggr\{{1\over\sin\theta}{d\over d\theta}
\left(\sin\theta{d\over d\theta}\right)&-&
\bigg({m^2+4-4m\cos\theta\over\sin\theta^2}\bigg)
\nonumber\\
&+&a^2\omega^2\cos^2\theta+4a\omega\cos\theta\Biggr\}
S_{-2,l,m}^{\omega}(\theta)=-E S_{-2,l,m}^{\omega}(\theta).
\end{eqnarray}
In the case $a\omega=0$ they reduce to the spin-weighted spherical harmonics,
i.e. $S_{-2,l,m}^{\omega}(\theta,a=0)={}_{-2}Y_{lm=0}(\theta)$.

\subsection{The boundary conditions for the Teukolsky equation}

The inhomogeneous radial equation can be solved with the physical 
requirement that gravitational waves must be purely ingoing at the black 
hole horizon, located at $r_+=M+\sqrt{M^2-a^2}$, and be purely outgoing at 
infinity. In mathematical language these statements translate into
\begin{equation}
R^H_{l,m\omega}(r^*) \sim \left\{
\begin{array}{ll}
f^2 e^{-i\omega r^*} & \ r \to r_+ ~(r^*\to-\infty) \\
{\cal Q}_T^{\rm in}\, (i\omega r)^{-1} e^{-i\omega r^*} 
+ {\cal Q}_T^{\rm out}\, (i\omega r)^3 e^{i\omega r^*} 
& \ r\to \infty ~(r^*\to\infty)
\end{array} \right. ,
\label{RH}
\end{equation}
 and
\begin{equation}
R^\infty_{l,m\omega}(r^*) \sim \left\{
\begin{array}{ll}
(i\omega r)^3 e^{i\omega r^*} & \ r \to \infty ~(r^*\to\infty)\\
{\cal P}_T^{\rm in}\, f^2 e^{-i\omega r^*} 
 + {\cal P}_T^{\rm out}\, e^{i\omega r^*} & \ r\to r_+ ~(r^*\to-\infty)
\end{array} \right.,
\label{rinfty}
\end{equation}
where ${\cal Q}_T^{\rm in}$ and ${\cal Q}_T^{\rm out}$, 
${\cal P}_T^{\rm in}$ and ${\cal P}_T^{\rm out}$ are constants
(independent of $r$) and $f(r)=\Delta/(r^2+a^2-am/\omega)$.
The coordinate $r^*$, as defined by Chandrasekhar is 
\begin{equation}\label{tortoise}
r^*_{\text ch}\equiv\int{dr\over f(r)}=r+M\ln\left(\frac{\Delta}{r_+^2}
\right)+\frac{M^2-am/(2\omega)}{\sqrt{M^2-a^2}}
\ln\left({r-r_+\over r-r_-}\right)\ ,
\end{equation}
where $r_\pm=M\pm\sqrt{M^2-a^2}$.
Note that this is {\it not} the standard tortoise coordinate, which is
instead defined by  $\int dr(r^2+a^2)/\Delta$.
The reason of our choice for this non-standard definition of 
$r^*$ (defined in Eq.\ (\ref{tortoise}))  
lies in the simplification of the forthcoming equations. From the numerical
point of view this choice should not be especially inconvenient.

The wronskian of these solutions is given by
\begin{equation}
W_{\rm T}(R^H,R^\infty) = \bigl( R_H R_\infty' - 
R_\infty R_H' \bigr)=-2 i \omega^3 {\cal Q}_T^{\rm in}\Delta.
\label{W_T}
\end{equation}
Note that the ``conserved Wronskian'' as considered in Refs.\ \cite{P97,D78}
is only the constant factor of $W_{\rm T}$, i.e.
$-2i\omega^3 {\cal Q}_T^{\rm in}$.

\section{CHANDRASEKHAR TRANSFORMATIONS}

It is called a Chandrasekhar transformation\cite{Ch75} the first
order differential operator that carries a solution to a Regge-Wheeler
like equation (with a real potential) into a solution to the homogeneous
Teukolsky equation. More explicitly

\begin{equation}
R^{H,\infty}(r)=\chi^{H,\infty}\, {\cal C} X^{H,\infty}(r),
\label{RCT}
\end{equation}
where $\chi^{H,\infty}$ are constants of normalization 
(see the Appendix) and
${\cal C}={\cal G}_{conv}(r)+{\cal G}_{div}(r){\cal L}$ with
${\cal L}=d/dr^*+i\omega$ is the first order operator. The form of
the functions ${\cal G}_{conv}(r)$ and ${\cal G}_{div}(r)$ (made explicit 
in the following subsection) are determined
by the requirement that the potential term, $V(r)$ of the resulting
Regger-Wheeler-like equation
\begin{equation}
\bar{\cal L}{\cal L}X=V(r)X,
\label{RWLE}
\end{equation}
be real. 

The asymptotic behavior of the solutions to this equation can be described
in terms of two solutions representing purely ingoing
gravitational radiation at the black-hole horizon, 
denoted by $X^H(r)$ and gravitational waves 
purely outgoing at infinity, denoted by
$X^\infty(r)$. Its asymptotic behavior can be characterized by

\begin{equation}\label{RHRW}
X^H_{l,m\omega}(r^*) \sim \left\{
\begin{array}{ll}
e^{-i\omega r^*} & \ r \to r_+ ~(r^*\to-\infty) \\
{\cal Q}_{RW}^{\rm in}\,e^{-i\omega r^*} 
+ {\cal Q}_{RW}^{\rm out}\, e^{i\omega r^*} 
& \ r\to \infty ~(r^*\to\infty)
\end{array} \right.,
\end{equation}
 and
\begin{equation}
X^\infty_{l,m\omega}(r^*) \sim \left\{
\begin{array}{ll}
e^{i\omega r^*} & \ r \to \infty ~(r^*\to\infty)\\
{\cal P}_{RW}^{\rm in}\, e^{-i\omega r^*} 
 + {\cal P}_{RW}^{\rm out}\, e^{i\omega r^*} & \ r\to r_+ ~(r^*\to-\infty)
\end{array} \right.,
\label{rinftyRW}
\end{equation}
where ${\cal Q}_{RW}^{\rm in}$ and ${\cal Q}_{RW}^{\rm out}$, 
${\cal P}_{RW}^{\rm in}$ and ${\cal P}_{RW}^{\rm out}$ are constants
(independent of $r$), which can be related to the corresponding 
Teukolsky constants, ${\cal Q}_T^{\rm in}$, ${\cal Q}_T^{\rm out}$, 
${\cal P}_T^{\rm in}$ and ${\cal P}_T^{\rm out}$.

The wronskian of these solutions is given by
\begin{equation}\label{WRW}
W_{\rm RW} \equiv
X^H\frac{dX^\infty}{dr^*}-X^\infty\frac{dX^H}{dr^*}
=2 i \omega {\cal Q}_{RW}^{\rm in}~.
\end{equation}

\subsection{Explicit form}

In 1975 Chandrasekhar\cite{Ch75} was able to reduce the Teukolsky's 
radial equation governing the general, non-axisymmetric, ingoing 
($s=+2$) gravitational perturbation of the Kerr black hole to a 
one-dimensional wave equation of the form\ (\ref{RWLE}). For the sake of 
future numerical convenience, we re-express, here, his results for the 
case of outgoing radiation ($s=-2$).
The functions that characterize the transformation\ (\ref{RCT}) are the
following 
\begin{equation}
\label{CNCT}
{\cal G}_{conv}=(i\omega)\sigma^{3/2}\left(V(r)
+b{\Delta^2\over\sigma^4}\right)\ ,\ \ 
{\cal G}_{div}=(i\omega)\sigma^{3/2}\left
({p/\sigma-k\Delta\over q-b\Delta}+2i\omega\right),
\end{equation}
where $f(r)=\Delta/(r^2+a^2-am/\omega)$.
$q$ and $p$ are functions of $r$, and $b$ and $k$ are constants, all
as defined below
\begin{eqnarray}\label{Fpq}
q(r)&\equiv&(\lambda-2)\sigma^2+3\sigma(r^2-a^2)
-3r^2\Delta\ ,\nonumber \\
p(r)&\equiv& q'\Delta-q\Delta'\ ,\\
b&\equiv&\pm3\left(a^2-am/\omega\right),
\nonumber \\ 
k&\equiv&\pm\left\{36M^2-2(\lambda-2)\left[(a^2-am/\omega)^2(5\lambda-4)
-12a^2\right]+2b\lambda(\lambda-2)\right\}^{-1/2}.\nonumber
\end{eqnarray}
The potential in the Regge-Wheeler equation is then given by 
\begin{equation}
V(r)=\Delta\left\{-b{\Delta\over\sigma^4}+{\lambda(\lambda-2)\over q+b\Delta}
-{(p/\sigma-k\Delta)\over (q+b\Delta)}{(kq-bp/\sigma)\over (q-b\Delta)^2}\right\}.
\label{NCP}
\end{equation}
Note that since $b$ and $k$ can be chosen with either sign, 
Eq.\ (\ref{NCP}) yields, in general, four different potentials. 
For the choice of the positive sign in the expression for $k$, 
one obtains two Zerilli-like potentials (depending on the sign of $b$),
which merge to the known Zerilli potential (for polar or even 
parity perturbations) in the limit of $a\to0$.
For the choice of the negative $k$, one obtains two possible 
Regge-Wheeler-like potentials, which reduces to the known Regge-Wheeler
potential (for axial or odd parity perturbation) in the limit of $a\to0$.
Although different, the four potential have the same asymptotic features
of vanishing at the horizon and of falling off as
${\cal O}(r^{-2})$ at infinity.

\section{A GENERAL PROCEDURE OF REGULARIZATION}

To regularize the solution of the radial Teukolsky equation with sources
that extend to infinity,
we can proceed along the same lines of Poisson\cite{P97} and generalize 
his procedure for the rotating case. 
It is also convenient to use here a notation parallel to the one used for 
the regularization of the Bardeen-Press equation. For instance, we can
express the radial source term (see Eq.\ (\ref{radialsource})) in the 
following form
\begin{equation}
{T(r)\over\Delta^2}=G g(r)e^{i\omega t(r)}~.
\label{R0} 
\end{equation}
An explicit form for $T(r)$ or equivalently for the constant $G$ and 
the function $g(r)$, is given in section V.
For the purpose of this section its sufficient to anticipate that
$\Delta^{-2}T(r)\sim {\cal O}(r^{-3/2})$ as $r\to\infty$.

The general solution to the inhomogeneous Teukolsky
equation can be expressed as
\begin{eqnarray}
R(r) &=& \frac{G\Delta}{W_{\rm T}} \Biggl\{ R^\infty(r) 
\biggl[
A + \int_c^r g(\tilde{r}) R^H(\tilde{r}) e^{i\omega t(\tilde{r})}
\, d\tilde{r}
\biggr] 
\nonumber \\ & & \mbox{}
+ R^H(r) \biggl[B + \int_r^d g(\tilde{r}) 
R^\infty(\tilde{r}) e^{i\omega t(\tilde{r})}\, d\tilde{r}
\biggr] \Biggr\}~,\label{3.18}
\end{eqnarray}
where $A$, $B$, $c$, and $d$ are constants to be determined
by the boundary conditions. 
Note that if we want to impose here the boundary conditions
(\ref{RH}) and (\ref{rinfty}), i.e. take $A=B=0$, $c=r_+$, and $d=\infty$,
the integrals do not converge when the superior limit goes to infinity.
We thus have first to regularize them for only then to impose
the boundary conditions. Such process of regularization leads to
Eq.\ (\ref{Resultado}) below.

We apply the Chandrasekhar transformation $\cal C$ given in 
Eq.\ (\ref{RCT}) to the following integrals
\begin{eqnarray}
\label{R1}
I^{H,\infty}
&&\equiv G\int{g(r) R^{H,\infty}(r) e^{i\omega t(r)} dr}
\nonumber\\ 
&&=\chi^{H,\infty}G\int{g(r) {\cal C} X^{H,\infty}
e^{i\omega t(r)} dr}\nonumber\\
&&=\chi^{H,\infty}G\int{g(r)\left({\cal G}_{\rm conv}
+ {\cal G}_{\rm div}{\cal L}\right)X^{H,\infty}
e^{i\omega t(r)} dr}\nonumber\\
&&=\chi^{H,\infty}G(I_{\rm conv} + I_{\rm div})~,
\end{eqnarray}
where
\begin{equation}
I_{\rm conv} = G\int{g(r){\cal G}_{\rm conv} e^{i\omega t(r)} 
X^{H,\infty}(r)dr},
\label{R2} 
\end{equation}
is a convergent integral since the integrand goes as ${\cal O}(r^{-1/2})$
for $r\to\infty$ 
(from Eq.\ (\ref{CNCT}), ${\cal G}_{\rm conv}\sim {\cal O}(r)$).
In this way we can restrict our attention to the divergent integral   
\begin{equation}
I_{\rm div}=G\int{g(r){\cal G}_{\rm div}
e^{i\omega t(r)} {\cal L} X^{H,\infty}(r)dr},
\label{R3}
\end{equation}
where now the integrand is ${\cal O}(r^{3/2})$ as $r\to\infty$
(from Eq.\ (\ref{CNCT}), ${\cal G}_{\rm div}\sim {\cal O}(r^3)$).

As noted by Poisson\cite{P97}, $I_{\rm div}$ can be regularized  
by introducing a function $h(r)$ and
integrating by parts i. e. writing it in the following form
\begin{eqnarray}
I_{\rm div} &=& \int\biggl[G g(r){\cal G}_{\rm div}
e^{i\omega t} {\cal L} X^{H,\infty} + \frac{d}{dr} 
\Bigl( h e^{i\omega t} {\cal L} X^{H,\infty}\Bigr) \biggr]\, dr
- h e^{i\omega t} {\cal L} X^{H,\infty} \biggr|_{\text{boundaries}}
\nonumber \\
&&\equiv\tilde{I}_{\rm div}- h e^{i\omega t} {\cal L} X^{H,\infty}
\biggr|_{\text{boundaries}},\label{R4}
\end{eqnarray}
where $h(r)$ is a function to be determined. 
Using the following property for the operator ${\cal L}$
\begin{equation}
{d\over dr}{\cal L}X={1\over f(r)}\left(V(r)+i\omega{\cal L}\right)X
\end{equation}
the new divergent integral becomes 
\begin{equation}
\tilde{I}_{\rm div} = 
\int{e^{i\omega t}
\left[\tilde\Gamma_{\rm div} {\cal L} X^A
+{h(r)\over f(r)}V(r)X^ {H,\infty}\right]dr}~,
\label{R5}
\end{equation}
where
\begin{equation}
\tilde\Gamma_{\rm div} = \frac{dh}{dr} + i\omega
\biggl( \frac{dt}{dr} + \frac{1}{f} \biggr) h
+G g(r){\cal G}_{\rm div}\ (=0)~.
\label{R6}
\end{equation}
Here $dt/dr$ is the equation of the trajectory, explicitly given in 
Section V, and the function $h(r)$ must be chosen so that 
$\tilde{I}_{\rm div}$ converges when $r\to\infty$. 
The natural and elegant choice of $\tilde\Gamma_{\rm div} = 0$ removes 
completely the divergence  in $\tilde{I}_{\rm div}$ as in the  
Schwarzschild case studied by Poisson\cite{P97},
\begin{equation}
h(r,a=0) = 8M^2\frac{1+x+2iM\omega x^3}{1+x},\quad
x=(2M/r)^{-1/2},
\end{equation}
but the power counting
shows that it is enough to remove all the dependence up to the first
power in $a$, i.e. terms like $(a/r)^2$ or higher produce convergent integrals.
Although is possible to solve formally the consequent first order
differential equation for $h(r)$, it's a very difficult task to find the
explicit form of the particular solution $h(r)$, with $a\neq 0$.
To handle this technical difficulty, one can,
for instance, numerically integrate this differential equation along
with the homogeneous Regge-Wheeler-like equation to obtain the final
solution\ (\ref{3.18}).

Using now Eqs.~(\ref{R1}), (\ref{R2}), (\ref{R4}), and (\ref{R5}) we
can express Eqs.~(\ref{R1}) in the following regularized form
\begin{eqnarray}
I^{H,\infty} &=& 
\chi^{H,\infty}\int{\left[G g(r){\cal G}_{\rm conv}
+{h(r)\over f(r)}V(r)\right]
X^{H,\infty}(r) e^{i\omega t(r)}dr}\nonumber\\
&-&\chi^{H,\infty} h(r) e^{i\omega t(r)} {\cal L} X^{H,\infty}(r)
\biggr|_{\text{boundaries}}~.\label{R8}
\end{eqnarray}

In summary, after the imposition of the boundary conditions 
\ (\ref{RH})-(\ref{rinfty}), the general solution \ (\ref{3.18})
to the radial Teukolsky equation can be written in the following 
regularized form 
\begin{eqnarray}\label{Resultado}
R(r) &=& \frac{G\Delta}{W_{rm T}} \Biggl\{\chi^{H} R^\infty(r) 
\biggl[
\int_{r_+}^r \left(G g(\tilde{r}){\cal G}_{\rm conv}
+{h(\tilde{r})\over f(\tilde{r})}V(\tilde{r})\right) X^H(\tilde{r})
e^{i\omega t(\tilde{r})}\, d\tilde{r}
\biggr]\label{3.8}\nonumber\\
&+& \chi^{\infty} R^H(r) \biggl[\int_r^\infty 
\left(G g(\tilde{r}){\cal G}_{\rm conv}
+{h(\tilde{r})\over f(\tilde{r})}V(\tilde{r})\right)
X^\infty(\tilde{r}) e^{i\omega t(\tilde{r})}\, d\tilde{r}
\biggr]\\
&+& \chi^H \chi^\infty{W_{\rm RW}\over 2} h(r) e^{i\omega t(r)}
{\cal G}_{\rm conv}\Biggr\},\nonumber
\end{eqnarray}
where the boundary term has been simplified 
by expressing $R^{H,\infty}(r)$ 
in terms of $X^{H,\infty}(r)$ by the transformation~(\ref{RCT}), and using
the following commutation relation $[{\cal L}, X^{H,\infty}]=
f(r)dX^{H,\infty}/dr$.

Note that the integrals are now not only convergent, but also expressed
in terms of the $X(r)$, solutions to the Regge-Wheeler-like equation, 
i.e. a one-dimensional equation with a real and short range potential.

To have an expression for the regularized integrals we
have to choose a Chandrasekhar transformation that gives an explicit
${\cal G}_{conv}$ and ${\cal G}_{div}$,
and produces a potential $V(r)$. The completion of the
solution is obtained by determining the normalization constants
$\chi^{H,\infty}$ that relate the behavior of the $R^{H,\infty}$ and the
$X^{H,\infty}$. In the Appendix we describe how to find them.

\section{THE SOURCE TERM}

In order to see explicitly how the regularization works, we must construct
the source term of the Teukolsky equation for the 
specific case of a particle of rest mass $m_0$ released from rest at
infinity and falling with zero angular momentum along the polar axis 
of a Kerr black hole of mass $M$ and angular momentum $J=aM$. 
Note that this choice is only dictated by the simplicity of the resulting
source term, but the procedure, as explained in Sec.\ IV, works for
any source extending to infinity.

The source term in the Teukolsky equation\ (\ref{master}),
$T=T(t,r,\theta,\varphi)$, can be constructed from the following
expression\cite{T73}
\begin{eqnarray}\label{fuente}
T&=&2\rho^{-4}\Biggr\{\left(-\hat{\Delta}+3\gamma-\bar{\gamma}+4\mu+\bar{\mu}
\right)\left[\left(-\bar{\hat{\delta}}-2\bar{\tau}+2\alpha\right)T_{n\bar{m}}
-\left(-\hat{\Delta}+2\gamma-2\bar{\gamma}+\bar{\mu}\right)T_{\bar{m}\bar{m}}
\right]\nonumber\\
&+&\left(-\bar{\hat{\delta}}-\bar{\tau}+5\bar{\beta}+7\alpha\right)\left[\left(
-\hat{\Delta}+2\gamma+2\bar{\mu}\right)T_{n\bar{m}}-\left(-\bar{\hat{\delta}}-
\bar{\tau}+2\bar{\beta}+2\alpha\right)T_{nn}\right]\Biggr\}~,
\end{eqnarray}
where the spin coefficients, operators (denoted by an overhat) 
and Kinnersley tetrad are explicitly given in Table\ \ref{tabla1}.

To begin with, we replace into it the following non-vanishing NP 
components of the stress-energy momentum tensor 
\begin{eqnarray}
&&T_{\bar{m}\bar{m}}\equiv T_{\mu\nu}\bar{m}^\mu\bar{m}^\nu 
= -{a^2\rho^2\sin^2\theta\over 2},
\nonumber \\
&&T_{n\bar{m}} \equiv T_{\mu\nu}n^\mu\bar{m}^\nu 
={ia\rho\sin\theta\over 2\sqrt{2}\Sigma}\left[T_{tt}(r^2+a^2)-
T_{rt}\Delta\right],
\label{S1} \\
&&T_{nn} \equiv T_{\mu\nu}n^\mu n^\nu 
={1\over 4\Sigma^2}\left[T_{tt}(r^2+a^2)^2+T_{rr}\Delta^2
-2T_{rt}(r^2+a^2)\Delta\right] . \nonumber
\end{eqnarray}

We then consider the stress-energy-momentum tensor for the trajectory 
specified above 
\begin{equation}
T^{\mu\nu}={m_0\over r^2}\tilde{f}(r)\delta(r-R(t))
\delta(\cos\theta-1)\delta(\varphi),
\label{S2}
\end{equation}
where $\tilde{f}(r)=\Delta/(r^2+a^2)$.

The only non vanishing components of the four-velocity are   
 \begin{eqnarray}
u^t&&\equiv {dt\over d\tau}= \tilde{f}(r)^{-1}\ ,\nonumber \\
u^r &&\equiv {dr\over d\tau}=-(1-\tilde{f}(r))^{-1/2}.\label{S4} 
\end{eqnarray}
The equation of the trajectory is 
\begin{equation}
{dt\over dr}=-\tilde{f}(r)^{-1}(1-\tilde{f}(r))^{-1/2}.
\label{trajectory}
\end{equation}

Finally, the source term of the radial equation\ (\ref{radial}) 
can be computed from the inverse Fourier transform of 
Eq.\ (\ref{separatedsource}),
\begin{equation}\label{radialsource}
T_{lm\omega}(r)=2\int{dt d\varphi d(\cos\theta) e^{i(\omega t-m\varphi)}
\Sigma T S_{-2,l,m}^{\omega}(\theta)}.
\end{equation}
Using symmetry arguments ($m=0$ for the particular trajectory 
we have considered here) we then obtain (in the notation of 
Eq.\ (\ref{R0}))
\begin{equation}
\hat{g}(r)=\frac{(r-ia)^2}{4r^2\Delta}\tilde{f}^{-1}
\left[(2-\tilde{f})(1-\tilde{f})^{-1/2}-2\right],
\label{gsombrero}\end{equation}
and
\begin{equation}
G=-16m_0N^0_{-2,l;0}(a\omega)~,
\label{G}\end{equation}
where the constant $N^m_{s,l;0}(a\omega)$ is defined by the
limiting behavior of spin weighted the spheroidal harmonics\cite{FHM88}
\begin{equation}
\lim_{\theta\to0}S_{sl}^m(\theta)=\theta^{|m+s|}N^m_{sl;0}(a\omega)\,.
\end{equation}
In the case $a\omega=0$ one has
\begin{equation}
G=-2m_0\sqrt{(l-1)l(l+1)(l+2)(2l+1)\over 4\pi}.
\end{equation}

\section{Summary and conclusion}

The procedure for regularizing the Teukolsky equation with source terms that
extend to infinity (infalling particle) in the frequency
domain was straightforward but lengthy. The finite result is given
in Eq.\ (\ref{Resultado}), where $G$, $W_T$, and $W_{RW}$ are given
in Eqs.\ (\ref{G}),\ (\ref{W_T}), and\ (\ref{WRW}) respectively,
and $\chi^{H,\infty}$ in the Appendix.
$R^{H,\infty}(r)$ and $X^{H,\infty}(r)$ are solutions to the radial
Teukolsky, Eq.\ (\ref{radial}) and Regge-Wheeler, Eq.\ (\ref{RWLE}),
homogeneous equations. ${\cal G}_{\rm conv}$ and ${\cal G}_{\rm conv}$
are given in\ (\ref{CNCT}) for the specific transformation chosen by
Chandrasekhar\cite{Ch75}. $\hat{g}(r)$ is given in Eq.\ (\ref{gsombrero}),
and bring all the dependence on the source, and hence on the particular
trajectory of the particle one wants to study. Finally, the regularizing
function $h(r)$ can be determined by solving the first order differential
equation\ (\ref{R6}). We repeat here that although we have studied 
explicitly only the case of the particle infalling from rest at infinity,
it is obvious that the regularization method applies to any type of
trajectory (the final expressions for will be more complicated though).

The method of regularization used in this paper has been chosen to parallel
that of Poisson\cite{P97}. It is clear that one could have used a different
procedure to regularize directly the Green function, as one does in field
theory. The final result, of course, has to be independent of the 
method chosen. It remains to be understood what is explicitly
the Green function of the radial Teukolsky equation in the case of
sources that extend to infinity. 
We would like to come back on this issue in the near future.

This long process to determine the regularized solution, although
valid in general, does not seem
very practical at the moment of finding a numerical solution to a
given astrophysical problem. In this sense, if one chooses to work in
the frequency domain, the approach by Nakamura and Sasaki\cite{SN82},
that find a Chandrasekhar transformation for both, the waveform and a
generic source that produce finite results, appears more appropriate.
In situations where one has to work in the time domain, like when
initial data are not negligible, one has to use the Teukolsky equation.
It is not yet known if in this case one has to perform again a
regularization of the results. Besides, one can actually implement
a procedure in order to obtain numerical results and compare them
with those obtained by solving the Sasaki-Nakamura equation (for the
particle infalling along the symmetry axis).
This will be the subject of research in our next paper.

\begin{acknowledgments}
C.O.L was supported by the NSF grant PHY-95-07719 and by research
funds of the University of Utah. The authors
thanks R. Price for useful comment and reading the original manuscript.
\end{acknowledgments}

\appendix

\section{Normalization constants}

To find the two normalization constants $\chi^H$ and $\chi^\infty$ we need
to look at the asymptotic behavior of the solutions to the wave equation
\ (\ref{RWLE}), given in Eqs.\ (\ref{RHRW})-(\ref{rinftyRW}). 
The asymptotic behavior at infinity do not
pose us any problem and the result of $\chi^\infty=1/4$, found by Poisson 
is valid here as well as in the $a\not=0$ case. This is so because the
effects of rotation on $R^\infty(r)$ vanish with a higher negative power
of $r$ than the corresponding nonrotational ones.

To study the asymptotic behavior of the solutions near the horizon
located at $r=r_+$, 
let $X(r)=Y(f)e^{-i\omega r^*}$. A simple change of variables 
tell us that $Y(f)$ satisfies a second order differential equation 
\begin{equation}
f^2f'^2Y_{,ff}+\left[f^2f''+ff'(f'-2i\omega)\right]Y_{,f}-V(f)Y=0.
\end{equation}

The solution to this differential equation can be given as a series
of the form $Y(f)=1+\sum_{n=1}^{\infty}a_nf^n$.
The first two coefficients are then found to be
\begin{eqnarray}
a_1&=&{V_{(1)}\over f'_{(0)}(f'_{(0)}-2i\omega)},\nonumber \\
a_2&=&{V_{(2)}-a_1\left[f''_{(0)}+2f'_{(1)}(f'_{(0)}-i\omega)-
V_{(1)}\right]\over 4f'_{(0)}(f'_{(0)}-i\omega)}~,
\end{eqnarray}
where we have used the notation $V(f)/f=V_{(1)}+V_{(2)}f+...$
($V_{(0)}=0$ since $V(r)$, given in Eq.\ (\ref{NCP}), vanishes 
at $r=r_+$).
And the derivatives of $f$ with respect to $r$ are
\begin{eqnarray}
f'_{(0)}&=&\frac{2\sqrt{M^2-a^2}}{\sigma_+},\nonumber \\
f''_{(0)}&=&
\frac{2}{\sigma_+}\left(1-\frac{4r_+\sqrt{M^2-a^2}}{\sigma_+}\right),\\
f'_{(1)}&=&\frac{f''_{(0)}}{f'_{(0)}}=
\frac{1}{\sqrt{M^2-a^2}}-\frac{4r_+}{\sigma_+},\nonumber
\end{eqnarray}
where the subindex ``+" means quantities evaluated at $r_+$.

We can now apply the Chandrasekhar transformation to $X^H$ and
determine the normalization constant $\chi^H$ such that
$R^H=\chi^H{\cal C}X^H$ reproduces the asymptotic
behavior\ (\ref{RH}). We then have asymptotically close to the horizon
\begin{eqnarray}
R^H&\sim& f^2e^{-i\omega r^*}=
\chi^H{\cal C}X^H=\chi^He^{-i\omega r^*}\Bigg\{
{\cal G}_{div}\frac{dY}{dr^*}+{\cal G}_{conv}Y\Bigg\}\nonumber\\
&=&\chi^He^{-i\omega r^*}\Bigg\{
\left({\cal G}_{div(0)}+{\cal G}_{div(1)}f+{\cal G}_{div(2)}f^2+{\cal O}(f^3)
\right)\left(a_1f+2a_2f^2+{\cal O}(f^3)\right)f'\nonumber\\
&&+\left({\cal G}_{conv(0)}+{\cal G}_{conv(1)}f+{\cal G}_{conv(2)}f^2
+{\cal O}(f^3)\right)\left(1+a_1f+a_2f^2+{\cal O}(f^3)\right)\Bigg\}.
\end{eqnarray}

Comparison order by order gives
\begin{equation}
{\cal G}_{conv(0)}=0\ ,\ \ \ \ a_1{\cal G}_{div(0)}f'_{(0)}+
{\cal G}_{conv(1)}=0\ ,
\end{equation}
on the horizon, and
\begin{equation}
\chi^H=\left\{{\cal G}_{conv(2)}+a_1\left[{\cal G}_{conv(1)}+
{\cal G}_{div(1)}f'_{(0)}+{\cal G}_{div(0)}f'_{(1)}\right]
+2a_2{\cal G}_{div(0)}f'_{(0)}\right\}^{-1}~.
\end{equation}

The explicit computation of the series developments of ${\cal G}$ and $V$
is tedious but straightforward. Direct use of Eq.\ (\ref{CNCT}) gives
\begin{eqnarray}
&&{\cal G}_{conv(1)}=i\omega\sigma_+^{3/2}V_{(1)}\ ,\ \ 
{\cal G}_{conv(2)}=i\omega\sigma_+^{1/2}\bigg\{{3\over2}
{\sigma'_+\over f'_{(0)}}V_{(1)}+\sigma V_{(2)}
+\frac{b}{\sigma_+}\bigg\}~,\\
&&{\cal G}_{div(0)}=i\omega\sigma_+^{1/2}\big\{2i\omega\sigma_+
-\Delta'_+\big\}\ ,\\ 
&&{\cal G}_{div(1)}=\frac{i\omega\sigma_+^{-1/2}}{2}
\bigg\{{(q_+-\sigma_+^3V_{(1)})\over\Delta'_+}+{3\sigma_+\over f'_{(0)}}
(2i\omega\sigma_+-\Delta'_+)\bigg\}~.
\end{eqnarray}

Now, from the expression of the potential\ (\ref{NCP}) we can obtain
\begin{eqnarray}
V_{(1)}&=&\frac{\sigma_+}{q_+}\left[\lambda(\lambda-2)+
\frac{\Delta_+'}{\sigma_+}\left(k+\frac{b\Delta_+}{\sigma_+}\right)\right]~,\\
V_{(2)}&=&\frac{\sigma'_+}{\Delta_+'}V_{(1)}-\frac{b}{\sigma_+^2}
-\lambda(\lambda-2)\frac{\sigma^2_+}{\Delta_+'}\frac{(q_++b\Delta_+)}{q_+^2}
\nonumber\\
&+&\frac{(k\sigma_+q_+-bp_+)(k\sigma_+\Delta_+'-p_+')-p_+(k\sigma'_+q_++
k\sigma_+q'_+-bp'_+)}{\Delta_+'q_+^3}\\
&+&\frac{p_+(k\sigma_+q_+-bp_+)}{\Delta_+'q_+^4}
\left(2\frac{\sigma'_+}{\sigma_+}q_++3q_+'-b\Delta_+'\right)
,\nonumber
\end{eqnarray}
where the subindex ``+" means quantities evaluated at $r_+$, and
$p=q'\Delta-q\Delta'$.

When $a=0$ expressions notably simplify and give
\begin{equation}
\chi^H=\frac{(1-2iM\omega)(1-4iM\omega)}
{iM\omega\bigr[(l-1)l(l+1)(l+2)-12iM\omega\bigr]}\ .
\end{equation}

\begin{table}
\caption{}
\begin{tabular}{llc}
Kinnersley tetrad&operators&spin coefficients\\
\tableline
\\
$n^\alpha=[r^2+a^2,-\Delta,0,a]/(2\Sigma)$&
$\hat{\Delta}=n^\alpha\partial_\alpha$&
$\rho=1/(r-ia\cos\theta)\ ,\ \ \ \ \ \ \ \ \ \ 
\tau=ia\rho\bar{\rho}\sin\theta/\sqrt{2}$\\
$m^\alpha=\bar{\rho}/\sqrt{2}[ia\sin\theta,0,1,i/\sin\theta]$&
$\hat{\delta}=m^\alpha\partial_\alpha$&
$\beta=\bar{\rho}\cot\theta/2/\sqrt{2}\ ,\ \ \ \ \ \ \ \ \ \ \ \ \ \ 
\mu=-\rho^2\bar{\rho}\Delta/2$\\
$l^\alpha=[(r^2+a^2)/\Delta,1,0,a/\Delta]$&
$\hat{D}=l^\alpha\partial_\alpha$&
$\alpha={\rho}(\cot\theta+2ia\rho\sin\theta)/2/\sqrt{2}$\ ,\ 
$\gamma=-\bar{\mu}-\rho\bar{\rho}\Delta'/4$\\
\end{tabular}
\label{tabla1}
\end{table}

\end{document}